\documentclass[10pt, conference, compsocconf]{IEEEtran}

\usepackage{cite}

\ifCLASSINFOpdf
   \usepackage[pdftex]{graphicx}

\else

\fi

\usepackage[cmex10]{amsmath}

\usepackage{url}

\usepackage{gensymb}

\usepackage[hidelinks]{hyperref}

\usepackage{cleveref}

\usepackage{pbalance}

\begin{document}

\title{Practical Non-Invasive Probing Attacks Against Novel Carbon-Nanotube-Based Physical Unclonable Functions}

\author{\IEEEauthorblockN{Nikolaos Athanasios Anagnostopoulos\IEEEauthorrefmark{1}, Alexander Braml\IEEEauthorrefmark{1}, Nico Mexis\IEEEauthorrefmark{1}, Florian Frank\IEEEauthorrefmark{1}, Simon Böttger\IEEEauthorrefmark{2}, Martin Hartmann\IEEEauthorrefmark{2},\\Sascha Hermann\IEEEauthorrefmark{2}\IEEEauthorrefmark{3}, Elif Bilge Kavun\IEEEauthorrefmark{1}, Stefan Katzenbeisser\IEEEauthorrefmark{1}, and Tolga Arul\IEEEauthorrefmark{1}\IEEEauthorrefmark{4}}
\IEEEauthorblockA{\IEEEauthorrefmark{1}Faculty of Computer Science and Mathematics, University of Passau, 94032 Passau, Germany\\
Emails: \{anagno02, braml11, mexis01, frank55, kavun01, katzen07, arul01\}@ads.uni-passau.de}%
\IEEEauthorblockA{\IEEEauthorrefmark{2}Research Center for Materials, Architectures and Integration of Nanomembranes,\\
and Center for Microtechnologies, Chemnitz University of Technology, 09126 Chemnitz, Germany\\
Emails: \{simon.boettger, martin.hartmann, sascha.hermann\}@zfm.tu-chemnitz.de}%
\IEEEauthorblockA{\IEEEauthorrefmark{3}Fraunhofer Institute for Electronic Nano Systems (ENAS), 09126 Chemnitz, Germany\\
Email: sascha.hermann@enas.fraunhofer.de}
\IEEEauthorblockA{\IEEEauthorrefmark{4}Department of Computer Science, Technical University of Darmstadt, 64289 Darmstadt, Germany\\
Email: arul@rbg.informatik.tu-darmstadt.de}%

}

\maketitle

\begin{abstract}

As the number of devices being interconnected increases, so does also the demand for (lightweight) security. To this end, Physical Unclonable Functions (PUFs) have been proposed as hardware primitives that can act as roots of trust and security. Recently, a new type of PUF based on Carbon NanoTubes (CNTs) has been proposed. At the same time, attacks and testing based on direct electrical probing appear to be moving towards non-invasive techniques. In this context, this work attempts to examine the potential for practical non-invasive probing attacks against the CNT-PUF, a novel PUF based on CNTs.
Our results indicate that direct probing might potentially compromise the security of this PUF. Nevertheless, we note that this holds true only in the case that the attacker can directly probe the wire corresponding to the secret value of each CNT-PUF cell. Thus, we can conclude that the examined CNT-PUFs are rather resilient to direct probing attacks, that non-invasive probing methods appear to be promising for testing such PUFs, and that, in order for the attacker to gain the full-length value of the secret, all the relevant channels would need to be probed. Nevertheless, as our work proves, practical non-invasive attacks against the CNT-PUF are feasible and adequate countermeasures need to be employed in order to address this issue.

\end{abstract}


\IEEEpeerreviewmaketitle

\section{Introduction}

As devices become more and more interconnected, especially in the context of the Internet of Things (IoT), the need for security keeps constantly increasing. To this end, hardware-based security primitives, such as Physical Unclonable Functions (PUFs~\footnote{The term ``Physical Unique Function'' can be used to more accurately describe a PUF.}) and True Random Number Generators (TRNGs), have been proposed as potential solutions that can act as security and trust anchors for the devices into which they are incorporated. 

In particular, PUFs have been proposed as a hardware-based security solution that allows for the secure generation and storage of cryptographic tokens, such as encryption keys, and thus for the implementation of a wide range of cryptographic protocols. PUFs are physical objects, such as hardware, that possess unique characteristics that are most often induced in the physical object by minor variations during its manufacturing process. Under specific conditions, e.g., when particular input is provided to a hardware device, to which we refer as the PUF challenge, the aforementioned PUF characteristics can be evaluated to acquire what is known as the PUF response, which in the case of hardware-based PUFs is most often binary logical values. Then, due to the inherent randomness, robustness, and uniqueness of each PUF instance, each of the corresponding PUF responses can be utilised for security applications.

Moreover, novel nanomaterials, such as Carbon NanoTubes (CNTs), graphene, and memristors, have recently started being utilised in the production of Integrated Circuits (ICs), not only in experimental chips intended only for research purposes, but even in commercially available products. At the same time, CNTs have recently been proposed for the implementation of PUFs~\cite{Hu2016,Burzuri2019}, because such PUFs may be more robust and tamper-resistant in comparison to ones based on silicon~\cite{Hu2016}, while additionally being compatible with the complementary metal-oxide-semiconductor (CMOS) fabrication process of electronic devices~\cite{9901422}. Such PUFs and their quality characteristics have already been examined in a number of recent works~\cite{7964978,doi:10.1021/acsaelm.9b00166,Hu2016,Burzuri2019,nano}.

Nevertheless, apart from the level of security that such PUFs can provide, i.e., the quality characteristics of their responses, also their resilience to attacks needs to be examined. To this end, in this work, we will examine whether non-invasive channel probing attacks can be performed against the CNT-PUFs proposed in~\cite{nano}. 

We need to note that, in general, probing is often considered as an invasive attack. However, recent research has shown that probing attacks can be performed in a non-invasive manner, most often utilising laser-based probing techniques~\cite{8167710,9027222}. Hence, in this work, we examine the potential for an attacker to acquire useful information on the CNT-PUF based on the leakage currents of the Carbon-NanoTube Field Effect Transistors (CNT-FETs) that form this PUF. For this reason, one might potentially also view our leakage-current probing as an attempt for a side-channel attack, to the extent that the leakage current is considered as a side channel.

In any case, however, our probing method is based on direct electrical probing of the leakage currents of the CNT-PUF, which would constitute a potential attack technique that would be extremely easy to perform. Additionally, to the extent that the relevant CNT-FETs used for the realisation of the CNT-PUF, are incorporated into a system as an IC component of its own, e.g., a dedicated CNT-PUF or a combination of a CNT-based sensor and a CNT-PUF, our probing method can actually be considered as rather practical, as long as it can be performed in a non-invasive manner, for example, by probing the relevant pins of the corresponding chip. In particular, only relatively low expertise and low to medium cost, related to the one-off purchase of electrical-measurement equipment, such as a Source-Measurement Unit (SMU), are required.

In this way, our work investigates the ability of an attacker to perform relatively low-cost and low-expertise non-invasive probing attacks in a rather practical manner. In particular, our work makes the following contributions:

\begin{enumerate}
    \item It measures the gate leakage current, $I_{G-leakage}$, of cells that correspond to the logical value 1 and of cells that correspond to the logical value 0, and examines whether an attacker can utilise this leakage current to gain information on the nature of a particular cell. Since a global gate is used in the CNT-PUF, i.e., the gates of all the cells are connected using the same wire, this current measurement does not allow the attacker to gain an advantage.
    \item It measures the combined leakage current at the drains of all cells found in columns other than the selected column and at the sources of all cells found in rows other than the selected row ($I_{leakage-non\_measured\_cells}$), which also, as expected, does not provide the attacker with an advantage, since the measured drain leakage currents tend to even out, as they are not dependent on the nature of the selected cell itself, and thus also not on the logical value to which it corresponds.
    \item It measures the drain leakage current, $I_{D-leakage}$, of a particular cell, which is rather strongly correlated with the nature of each cell, and thus also with the logical value to which each cell corresponds. Thus, in this case, it is shown that an attacker can gain a significant advantage and practically guess the logical value of the corresponding cell with extremely high~probability.
    \item It discusses the presented results and their significance for the security of the CNT-PUF, especially in the context of the concatenation of the drain currents, $I_{D}$, of all CNT-PUF cells forming the secret of this PUF, i.e., its response.    
\end{enumerate}

The rest of this work is organised as follows.~\Cref{sec:background} presents background information relevant to PUFs in general, and the CNT-PUF in particular.~\Cref{sec:rw} briefly explores works related to the two so-far distinct topics of IC probing and CNT-based PUFs. Then,~\Cref{sec:attacks} examines the potential for attacks against the CNT-PUFs based on non-invasive probing. Finally,~\Cref{sec:conclusion} summarises our results and propose potential future research topics, in order to conclude this work.

\section{Background Information on the CNT-PUF}
\label{sec:background}

As already mentioned, the specific conditions as well as any input or stimulus used in order to measure a PUF are referred to as the PUF response, while the output of the PUF, e.g., the concatenation of the values of its cells, is referred to as the PUF response.

In general, the quality of a PUF as a security mechanism is highly dependent on the existence of Challenge-Response Pairs (CRPs) that are unique per device and rather unpredictable. Thus, in order to evaluate the quality of the PUF responses produced, the well-known metrics of the Shannon entropy, the Hamming weight, and the Hamming distance are often employed. 

The CNT-PUF response is based on the characterisation, through a threshold $I_D$ value, of the cells of a monolithic 12$\times$12 crossbar array of CNT-FETs, as either conductive (acting either as true conductors or as semiconductors) or non-conductive (acting as insulators). The cells belonging to the former category are assigned the logical value of $\mathtt{1}$, and the cells belonging to the latter category, the logical value of $\mathtt{0}$, leading to a 144-bit PUF response that has proven to be extremely stable~\cite{nano}.

In particular, the drain current $I_D$ of each CNT-FET is measured under the influence of a gate-source voltage $V_{GS}$ equal to $-2.5$V and a drain-source voltage $V_{DS}$ equal to $-1$V. Nevertheless, it has been observed that the measurements of $I_D$ inherently incorporate some noise, leading to slightly different values for the same cell, even for the exact same value of $V_{GS}$. However, as such measurement values tend to concentrate within a limited region, i.e., to form a cluster, each cell is assigned a logical value based on whether the measured $I_D$ value is above or below a threshold value that is common for all cell measurements. Essentially, cells providing a high $I_D$, whose CNTs act as true conductors or semiconductors, are assigned the logical value of 1, and cells providing a low $I_D$, whose CNTs act as insulators, are assigned the logical value of 0.

The concatenation of the logical values assigned to all the cells forms the response of this PUF, with the provided $V_{GS}$ and $V_{DS}$, and the order in which the CNT-FET cells are measured, as well as other relevant conditions, such as the ambient temperature, forming the relevant PUF challenge. As it is evident, this method allows only for the production of a single 144-bit binary response for each crossbar array.

\section{Related Work}
\label{sec:rw}

In general, there appear to be no works relevant to testing in depth the security of PUFs based on CNTs by physically probing them so far, while it has also been claimed that micro-probing could easily break down a PUF based on CNTs and, thus, destroy its information~\cite{8067537}. Thus, we will briefly examine here the two so-far distinct topics of CNT-based PUFs and physical probing. 

As already mentioned, CNT-based PUFs and their quality characteristics have been examined in a number of recent works. In particular, rather robust CNT-based PUFs have been proposed in the literature, but their security has only minimally been examined.

In 2016, Hu \textit{et al.}~\cite{Hu2016} investigated the ability of self-assembled CNTs arranged in a $64\times40$ crossbar structure to serve as either a binary or a ternary PUF. For both cases, a maximum fractional intra-device Hamming distance of $\approx10\%$ is reported. Additionally, the work of Hu \textit{et al.} examines the stability of the relevant CNTs through measurements at $25${\degree}C and at $85${\degree}C. Thus, the security that such PUFs provide is rather well-tested, but their own security is only briefly examined in this work.

In 2017, Moradi \textit{et al.}~\cite{7964978} proposed novel CNT-based PUF types utilising either the voltage or the current output of CNT-FETs. This work provided simulation results for the temperature range between $0${\degree}C and $100${\degree}C, demonstrating reliability values that would correspond to fractional intra-device Hamming distances of at least $3.33\%$ for the voltage-based PUF and of at least $12.72\%$ for the current-based PUF. Thus, the security that these PUFs provide is rather proven, but their resilience to attacks is not considered.

In 2018, Liu \textit{et al.}~\cite{8067537} discussed the combination of a CNT-FET crossbar structure with the Lorenz chaotic system, in order to provide a PUF that would be resistant to machine-learning attacks. Only simulation results are provided and the work focuses on the uniqueness, randomness, and unpredictability of the PUF responses, with no results being reported regarding the stability of these responses. The authors consider machine-learning attacks, namely attacks based on the employment of Support Vector Machines (SVMs), Deep Belief Networks (DBNs), Linear Regression (LR), and Evolution Strategies (ES), performed both against the CNT-based PUF on its own and its combination with the Lorenz system of equations. In general, the authors rather prove that the Lorenz chaotic system cannot be machine-learned using the relevant methods tested, and thus its combination with the CNT-based PUF also cannot be machine-learned. 

However, it is worth noting that the CNT-based PUF tested is a {\emph weak} PUF, i.e., it has ideally only one CRP, and thus should always be machine-learned with 100\% accuracy. Nevertheless, the authors of~\cite{8067537} report that this is not the case for all the machine-learning methods tested. This could be attributed to the simulated CNT-based PUF responses being noisy, as would be the case in reality, but the authors do not clarify if that is indeed the case. 

It is also rather notable that it is claimed in~\cite{8067537} that micro-probing could easily destroy a CNT-based PUF; a claim that is not truly explored and which would only be relevant to invasive probing. Another bold claim of this work is that CNT-based PUFs are difficult to clone and, therefore, resistant to physical attacks. Since the response of CNT-based PUFs is dependent only on whether each relevant PUF cell is conductive or insulative, this claim is rather easy to disprove. In this work, we prove that a CNT-PUF cell's state can be measured through the drain leakage current, i.e., even when the cell is not triggered with a gate-source voltage difference, and thus can potentially be easily cloned, as long as an attacker has physical access to the relevant wire(s) of the device.

A work by Moon \textit{et al.}~\cite{doi:10.1021/acsaelm.9b00166} regarding PUFs produced by all-printed CNT networks, which was published in 2019, demonstrated no significant changes regarding the reliability of the fabricated CNTs over 10,000 measurement cycles and 14 days. However, the relevant resistance characteristic of individual CNTs was reported to have changed up to $16.7\%$, and results for temperature variations between $25${\degree}C and $80${\degree}C exhibited an average difference of up to $30\%$. Moreover, the authors proved that this PUF is robust and stable against tampering attacks such as high-temperature baking, light illumination, and radiation exposure. Nevertheless, the authors do not examine direct electrical probing of the PUF, and note that a local physical attack would destroy the device without gaining access to the secret, rather referring to an invasive attack. On the contrary, our work shows that a non-invasive probing attack can potentially access the device's secret, as long as the relevant wires and signals are not adequately protected and/or monitored.

Also in 2019, Burzur{\'i} \textit{et al.}~\cite{Burzuri2019} examined the ability of single-walled carbon nanotubes to be used for the creation of PUFs. This work demonstrated good results regarding the uniqueness and the robustness of the fabricated PUFs, reporting fractional intra-device Hamming distances of $6.3\%$ after two weeks and $8.3\%$ after two months. However, this work also does not consider the resilience of the relevant PUF to attacks.

Finally, most recently, another work by Böttger \textit{et al.}~\cite{nano} proposed the CNT-PUF, a highly robust PUF, whose response is, however, not fully stable. The authors propose identifying the few unstable bits during enrollment and excluding them from the production of future responses, which almost always results in robust responses thereafter. An average fractional intra-device Hamming distance of only $1\%$ is reported for the $I_{D}$ threshold that provides the highest robustness, while an average fractional intra-device Hamming distance of $3\%$ is reported for other less optimal values of such a threshold value. Again, however, no attacks against this PUF are considered.

When probing methods are considered, we can distinguish between invasive and non-invasive methods. Regarding the former type of probing, the relevant literature, e.g.,~\cite{8067537}, appears to suggest that its employment in the context of CNT-PUFs would not be effective and/or the relevant probing attempt would be detectable. Hence, non-invasive probing would be the preferable way of probing these PUFs, both in the context of an attack as well as for testing purposes. 

To this end, we have already noted that non-invasive probing is most often realised using laser-based probing techniques~\cite{8167710,9027222}. Nevertheless, depending on the case, other non-invasive probing techniques may be possible. For example, in 2010, Fuhrmann \textit{et al.}~\cite{10.1063/1.3373415} utilised surface acoustic waves as probing means to measure in a non-invasive manner the persistent photoconductivity of a ZnCdSe/ZnSe quantum-well heterostructure on a LiNbO$_3$~substrate. 

Moreover, in 2017, Sugawara \textit{et al.}~\cite{8167710} proposed utilising a detection technique based on non-invasive laser probing as a side channel both to the advantage of a malicious attacker as well as for testing and fault analysis. More information on laser fault injection and countermeasures can be found in~\cite{mastersthesis} and other works.

More recently, in  a work from 2019, Rahman and Asadizanjani~\cite{9027222} assessed back-side attacks against instances of System-on-a-Chip (SoC) technology, including electro-optical probing, laser fault injection, and other similar techniques, as well as relevant back-side testing methods. The authors noted that such semi-invasive and non-invasive probing techniques allow for run-time testing from the back side of a chip, which is most often left unprotected. The authors also classify the different adversaries into a number of categories, based on their having access to the chip design itself or not, and on their expertise.

Finally, in 2022, Sakamaki \textit{et al.}~\cite{9896502} demonstrated the non-invasive probing measurement of transmission lines on CMOS chips from 100 MHz to 500 GHz. The authors managed to reduce probe skating down to 10$\mu m$ using a precision-controlled probe station, which allowed for the use of extremely small contact pads, and helped preserve the characteristics of the contact material even after repeated probe touchdowns that normally would have worn out the pads. The authors suggest that non-invasive probing can be used to characterise Complementary Metal–Oxide–Semiconductor (CMOS) passive devices, which do not require Direct Current (DC) biasing.

Generally, different invasive probing techniques have been examined rather thoroughly in a number of works. For example, in 1999, Kuhn and Kömmerling~\cite{KUHN199928} examined the physical security of smartcards, including tampering methods, such as invasive micro-probing. It is important to note that this work examined also the exact costs of such attacks; an aspect of (physical and/or other) attacks that is most often neglected. This work was an extension of the one by Kömmerling and Kuhn~\cite{271706}, where the relevant topics had been first discussed.

In 2011, Skorobogatov~\cite{Skorobogatov2012} broadly examined physical attacks and tamper resistance, focusing, among others, on invasive micro-probing, while also discussing optical probing and laser-based techniques.

Moreover, in 2013, Helfmeier \textit{et al.}~\cite{10.1145/2508859.2516717} examined back-side invasive micro-probing attacks for invasive IC analysis, editing, and other similar applications.

In 2017, Shi \textit{et al.}~\cite{Shi2017} reviewed invasive probing attacks and studied potential countermeasures and designs against~them.

Finally, in 2018, Rahman \textit{et al.}~\cite{8494856} had a work published on physical inspection and attacks, which, among others, discusses both electrical and optical probing, considering the former as an invasive method and the latter as a non-invasive~technique.

In any case, we note that none of these works breached, let alone addressed, the subject of probing CNT-based PUFs, especially in a non-invasive manner.

\section{Practical Non-Invasive Probing Attacks Against CNT-PUFs}
\label{sec:attacks}

In this section, we first consider an attacker model relevant to non-invasive probing attacks in the context of CNT-PUFs, then, discuss the regular way in which such PUFs are measured, and, subsequently proceed to present potential non-invasive attacks against such PUFs, discuss the relevant results and their significance, as well as possible countermeasures against such attacks. 

\begin{figure*}[!t]
\centering
\includegraphics[width=\linewidth]{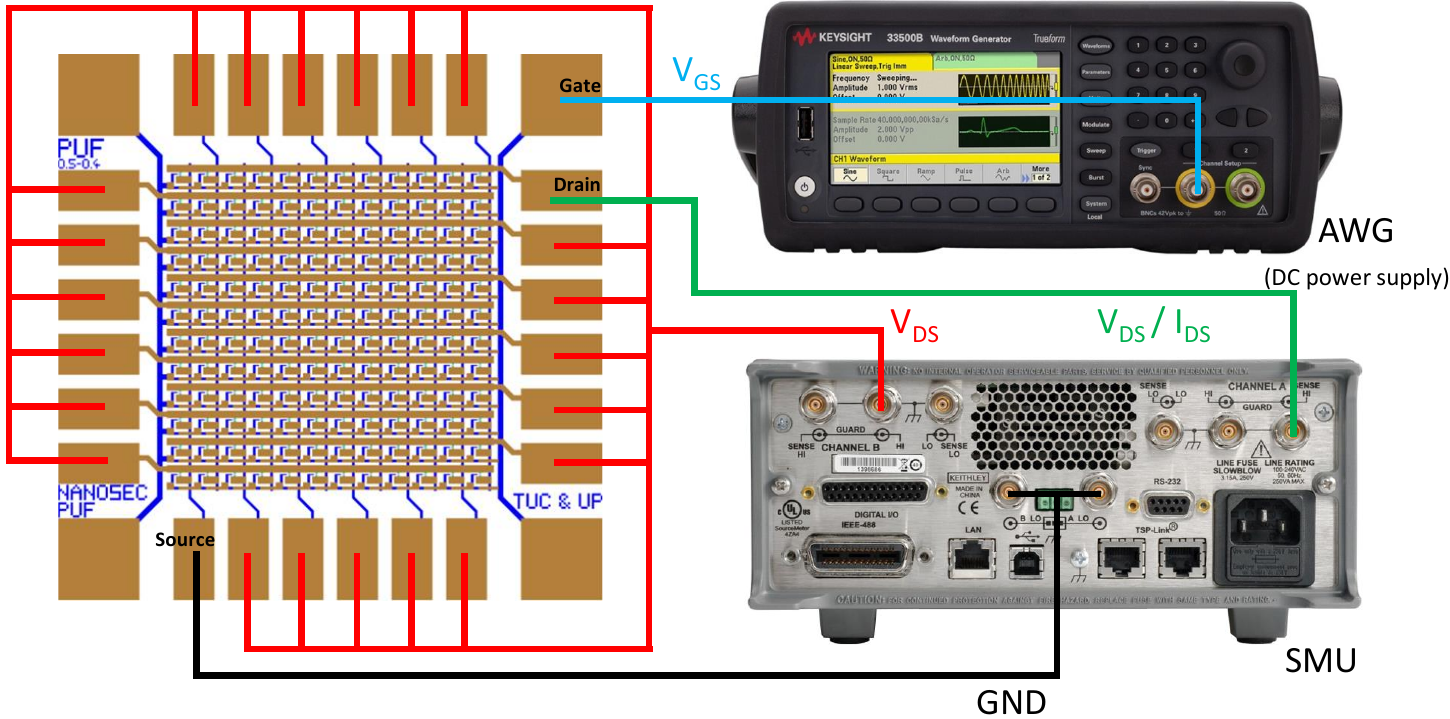}
\vspace{-20pt}
\caption{Measurement setup. Here, an Arbitrary Waveform Generator (AWG) is used to provide the gate voltage, instead of a DC power supply, as it allows for faster testing.}
\vspace{-20pt}
\label{fig:measurement}
\end{figure*}

\begin{figure}[!t]
\centering
\includegraphics[width=0.475\linewidth]{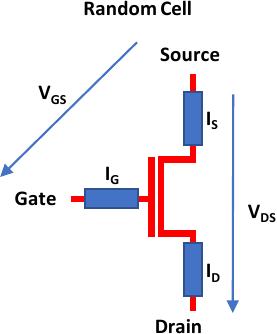}
\vspace{-5pt}
\caption{The layout of a CNT-FET, forming a CNT-PUF cell.}
\vspace{-20pt}
\label{fig:random}
\end{figure}

\subsection{Attacker Model}

In this work, we consider an attacker with physical access to the device containing the CNT-PUF. The attacker can probe wires relevant to the CNT-PUF in a non-invasive way, in order to provide DC voltage and measure the relevant electrical currents. We assume that the CNT-PUF may be a standalone chip, which may also potentially be incorporated into a larger system. Then, the attacker may directly probe the pins of the CNT-PUF chip, or any wires connected to them, in a non-invasive way, as long as these are kept accessible or can be accessed in such a way, e.g., through laser-based probing.  

The attacker would need to have access to relevant electrical-measurement equipment, including an SMU, as the current levels are extremely small, of an order of $10^{-12}A$, i.e., some picoamperes, in the worst case. Additionally, a DC power supply of regular (not necessarily high) precision may also be required. The relevant cost for the one-off purchase of such equipment may be some thousands of euros or, at most, tens of thousands of euros. Thus, our attack may be considered as a low-to-medium-cost attack, which however can be performed by an individual. Such an individual needs to possess some information regarding the way in which the CNT-PUF works, which, however, is publicly available and does not really require the expertise of a specialist. That means that a person of low expertise, with basic knowledge in the field of electricity can perform the attacks described in this work. In particular, the most challenging issue that the attacker will need to address is deciding for every type of leakage-current measurement on an adequate threshold in order to separate between currents corresponding to conductive and non-conductive cells. This decision may require some experience, which the attacker is bound to quickly acquire (even on-the-field) after a number of cells have been measured. To this end, experience may also be acquired by measuring an own CNT-PUF device, before the attack is performed on the target CNT-PUF device; nevertheless, this is not truly necessary for the attack to be successful.

Thus, the attacker may be a person of low expertise with access to medium-cost electrical-measurement equipment, which he/she does not need to own, and with physical access to the CNT-PUF device.

In general, as our probing attacks are based on direct electrical probing of the leakage currents of the CNT-PUF, we believe that they constitute practical attacks that would be extremely easy to perform and which also a rather small amount of time to perform, in the order of minutes.

\begin{figure*}[!t]
\centering
\includegraphics[width=0.75\linewidth]{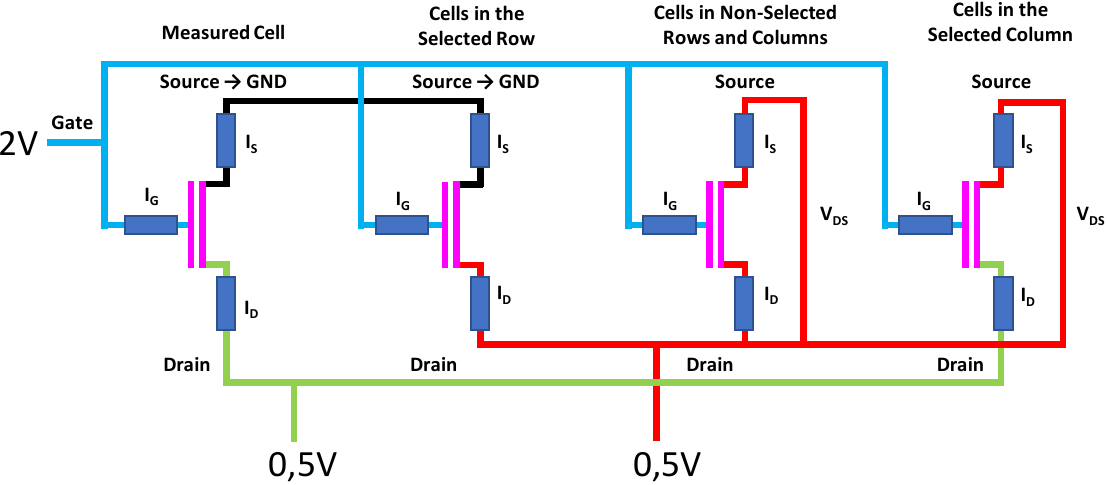}
\vspace{-7.5pt}
\caption{The  equivalent cell layout of the employed measurement setup.}
\vspace{-15pt}
\label{fig:measuredlayout}
\end{figure*}

\subsection{Regular Measurement Procedure}

It is important to first examine the regular way in which the CNT-PUF structure may be measured, in order to provide the reader with some insights into the way the proposed attacks may work out and why we consider them as practical.

In order to measure each cell of a CNT-PUF, the relevant row and column are selected, and the source of the cell is connected to the ground, while a low level of voltage (typically $0.5$V or $1$V) is provided to the drain. Without loss of generality, also $-0.5$V or $-1$V may be provided to the drain. The sources and the drains of all the cells found in rows and columns that have both not been selected are provided with the same level of voltage as the drain of the selected cell, in order to exactly avoid current leakage and other parasitic phenomena among the different rows and columns. Then, a certain voltage level is provided to the global gate, typically of $2$V or $2.5$V, which would lead to all the cells being turned on~\footnote{This is exactly the reason why the sources and the drains of the cells that are in rows and columns that have both not been selected should be provided with the same voltage, in order to avoid that these cells conduct; in this way, we ensure that these cells have the least possible effect on the measurement. Unavoidably, however, while all the cells in the selected row other than the selected cell will also not conduct, as both their sources and drains are at the same voltage level, all the cells in the relevant selected column will conduct, as their sources are grounded and their drains are at the voltage level of the drain of the cell to be measured.}. Again, without loss of generality, also $-2.5$V or $-2$V may be provided to the gate. At this point, the drain current of the selected cell is measured, and based on its value the cell can be classified as conducting or non-conducting, and thus be assigned the logical value of 1 or 0, respectively.

The measurement setup can be seen in~\Cref{fig:measurement}. There, an Arbitrary Waveform Generator (AWG) provides the gate voltage, for faster testing. The layout of a CNT-FET, which forms a CNT-PUF cell is shown in~\Cref{fig:random}. Hence, the equivalent cell layout of the employed measurement setup is illustrated in~\Cref{fig:measuredlayout}.

\begin{figure*}[!t]
\centering
\includegraphics[width=0.75\linewidth]{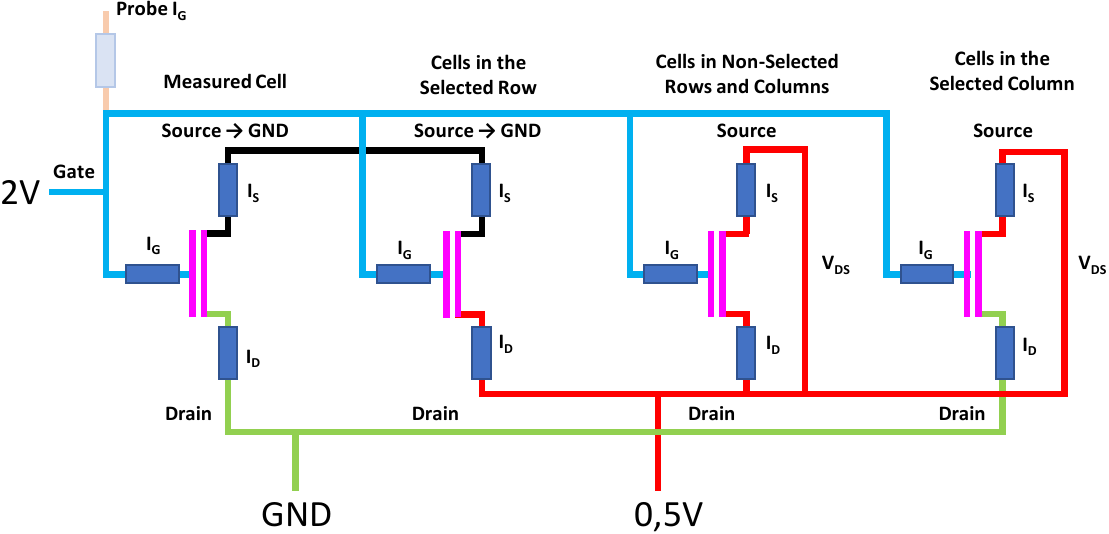}
\vspace{-7.5pt}
\caption{The  equivalent cell layout of the first measurement method: Measuring the gate leakage current of cells selected for measurement through a non-invasive probe.}
\vspace{-10pt}
\label{fig:onelayout}
\end{figure*}

\begin{figure*}[!t]
\centering
\includegraphics[width=0.6\linewidth]{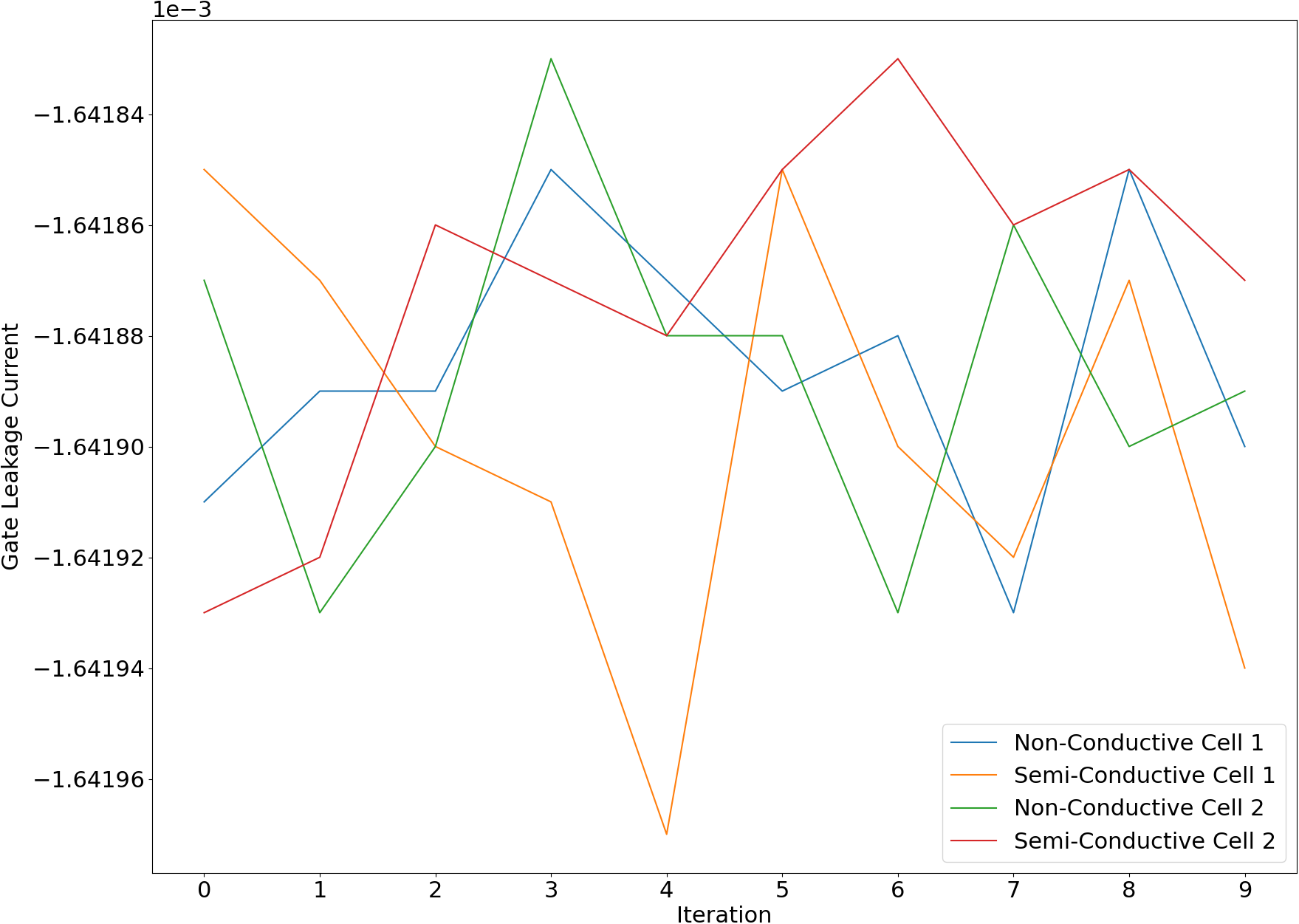}
\vspace{-7.5pt}
\caption{Ten measured $I_{G-leakage}$ values for each of two CNT-PUF cells corresponding to the logical value of 1 (red and orange) and each of two cells corresponding to the logical value of 0 (blue and green). It is clear that, in this case, the nature of the cells cannot be distinguished using these measurements.}
\vspace{-15pt}
\label{fig:igleakage}
\end{figure*}

\subsection{Examined Non-Invasive Probing Attacks}

As already described, we test three different probing measurements:

\subsubsection{Measuring the Gate Leakage Current of the Cells Selected} 

We measure the gate leakage current, $I_{G-leakage}$, of cells selected for measurement, by each time setting the relevant cell's $V_{DS}=0$, by grounding both its source and its drain. We also set $V_{GS}=2$, while the sources and the drains of the cells in rows and columns not selected are provided with $0.5$V, leading these cells to also exhibit $V_{DS}=0$. Then, we measure the overall $I_{G}$ for the global gate, which, however, corresponds to the gate leakage of not only that cell, but also of all the cells in the rows and columns that have both not been selected. The latter cells also have some voltage provided to their sources and drains, while the drain and the source of the selected cell are grounded. At the same time, the cells in the same row as the selected cell but on non-selected columns are turned on, with $V_{DS}=0.5$, while the cells in the same column as the selected cell but on non-selected rows are turned on, with $V_{DS}=-0.5$. Due to the crossbar structure of the CNT-PUF, and the use of a global gate, it is impossible to measure the gate leakage of only a single cell. However, in the future, the sources of the cells in non-selected rows could also be grounded, in order to have most of the cells turned on, and only the cells in the selected column not turned on. The equivalent cell layout for this probing measurement is shown in~\Cref{fig:onelayout}.~\Cref{fig:igleakage} illustrates the results relevant to such probing measurements for two semiconducting and two non-conducting cells~\footnote{For reasons of brevity and clarity, but without loss of generality, we present results for only four cells, two semi-conducting ones and two non-conducting. Moreover, we chose to present results for only semi-conducting and non-conducting cells, as the characteristics of semi-conducting cells are rather closer to those of non-conducting cells than the characteristics of fully conducting cells would have been.\label{foot3}}. As it is clear, there is no way to distinguish between semi-conducting and non-conducting cells using this method.

\subsubsection{Measuring the Leakage Current of the Cells Found in Rows and Columns that Have Not Been Selected}

\begin{figure*}[!t]
\centering
\includegraphics[width=0.75\linewidth]{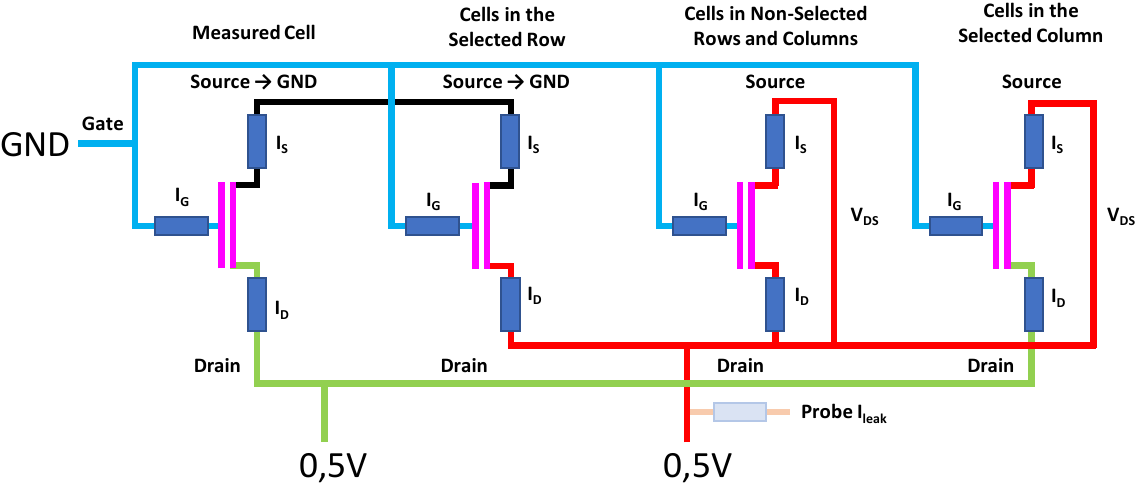}
\vspace{-7.5pt}
\caption{The  equivalent cell layout of the second measurement method: Measuring the leakage current of the cells found in rows and columns that have not been selected.}
\vspace{-10pt}
\label{fig:twolayout}
\end{figure*}

\begin{figure*}[!t]
\centering
\includegraphics[width=0.6\linewidth]{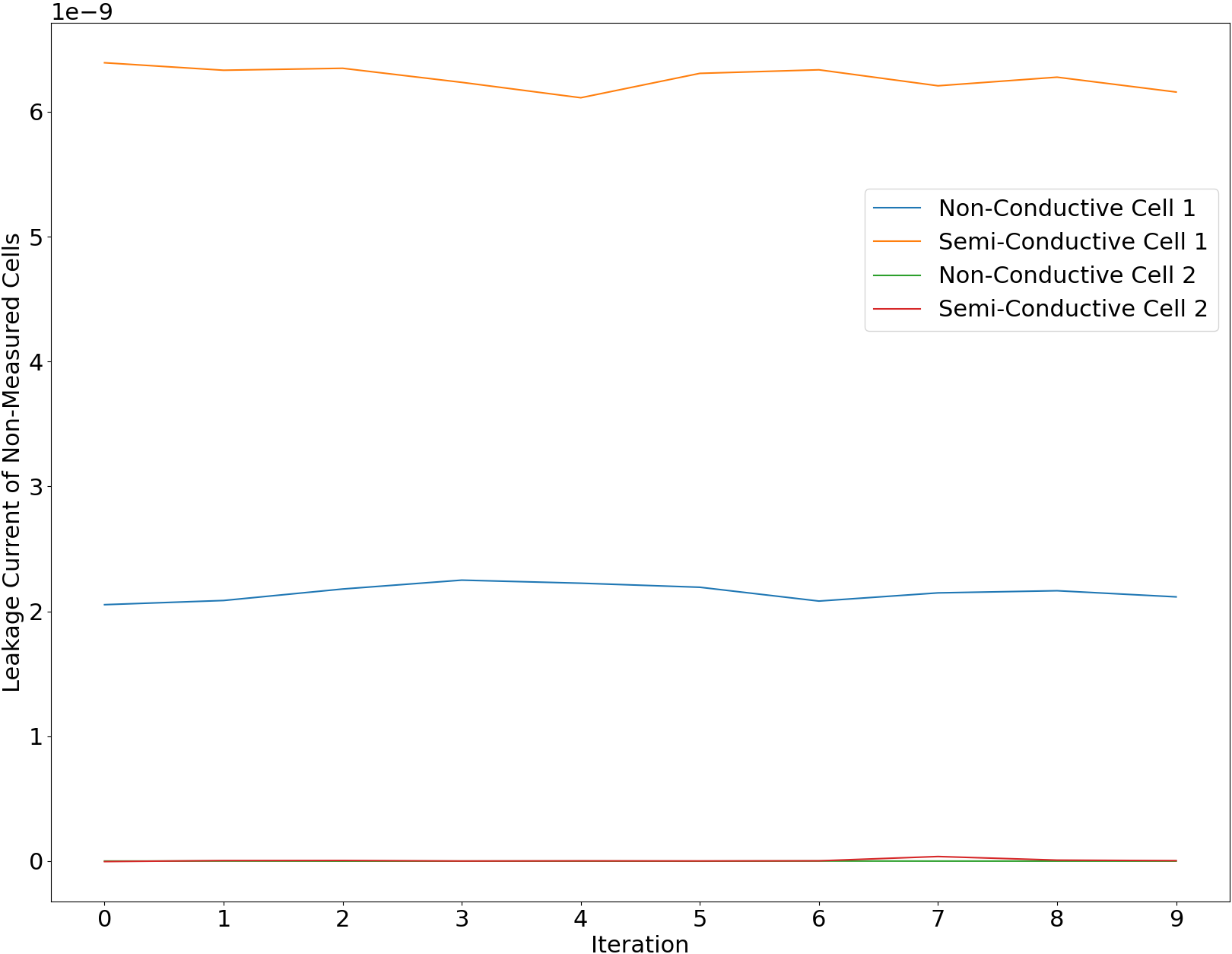}
\vspace{-7.5pt}
\caption{Ten measured $I_{leakage-non\_measured\_cells}$ values for each of two CNT-PUF cells corresponding to the logical value of 1 (red and orange) and each of two cells corresponding to the logical value of 0 (blue and green). It is clear that, also in this case, the nature of the cells cannot be distinguished using these measurements.}
\vspace{-10pt}
\label{fig:idothersleakage}
\end{figure*}

In the second probing measurement method, we utilise a probe to measure the combined leakage current at the drains of all cells found in columns other than the selected column and at the sources of all cells found in rows other than the selected row ($I_{leakage-non\_measured\_cells}$). To this end, we ground the gate, and provide $0.5$V to the sources and the drains of the cells in rows and columns not selected, leading these cells to exhibit $V_{DS}=0$. At the same time, we ground the source of the selected cell, which would normally be measured, and provide $0.5$V to its drain. This means that all the cells in the selected row have $V_{DS}=0.5$, while both the sources and the drains of the rest of the cells are provided with $0.5$V, so that they exhibit $V_{DS}=0$. As the global gate is grounded, none of the cells is expected to be turned on. The equivalent cell layout for this probing measurement is shown in~\Cref{fig:twolayout}.~\Cref{fig:idothersleakage} illustrates the results relevant to such probing measurements for two semiconducting and two non-conducting cells~\textsuperscript{\ref{foot3}}. It is clear that, again, there is no way to distinguish between semi-conducting and non-conducting cells using this method.

\begin{figure*}[!t]
\centering
\includegraphics[width=0.75\linewidth]{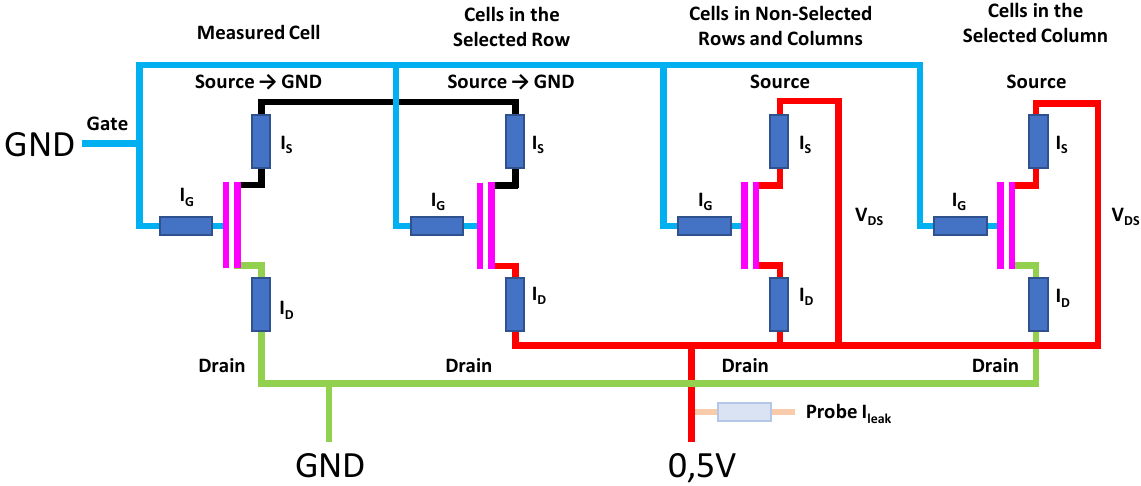}
\vspace{-7.5pt}
\caption{The  equivalent cell layout of the modified second measurement method: Measuring the leakage current of the cells found in rows and columns that have not been selected, with the drain of the selected cell also being grounded.}
\label{fig:twolayoutIDS}
\end{figure*}

\begin{figure*}[!t]
\centering
\includegraphics[width=0.6\linewidth]{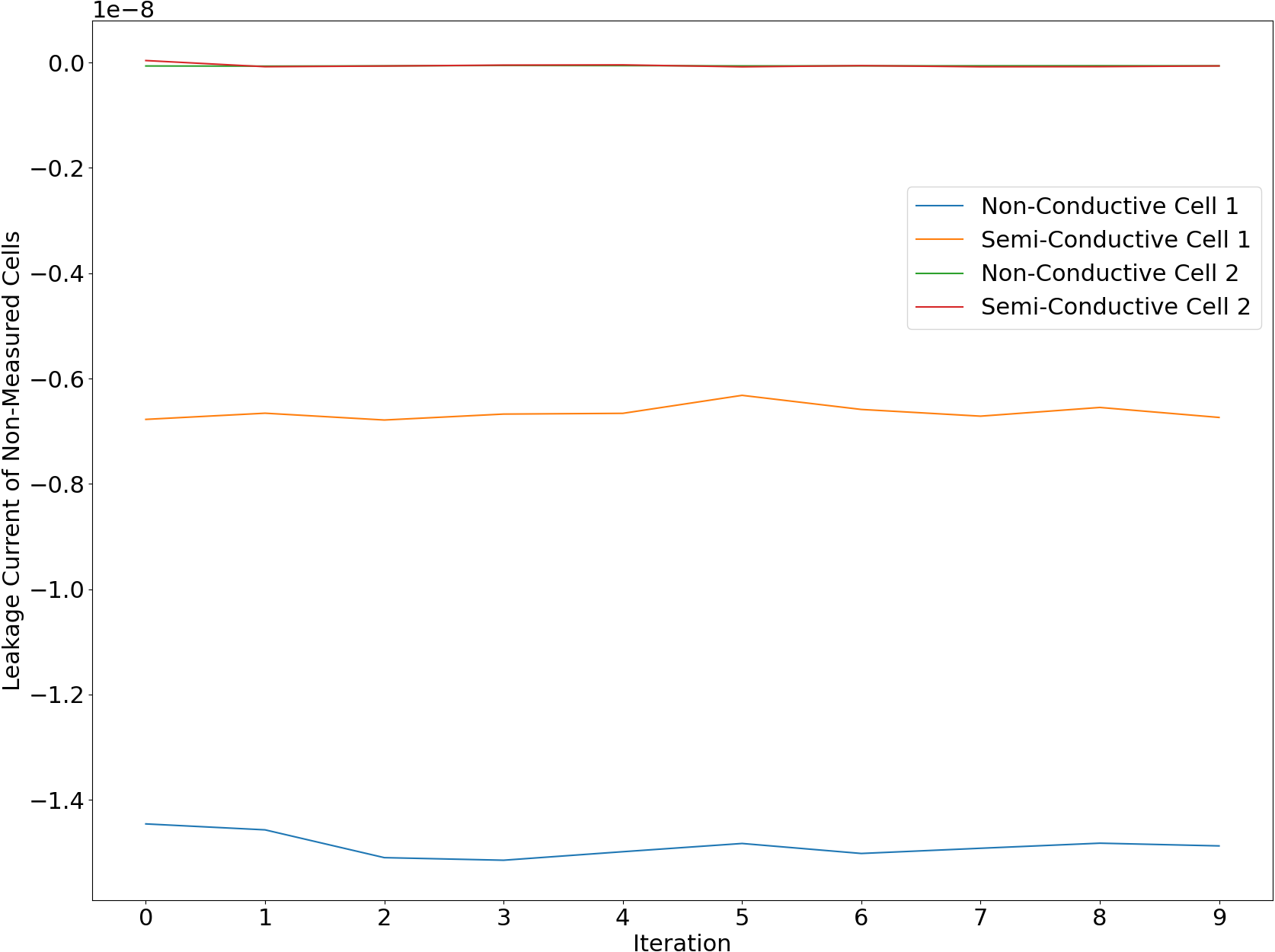}
\vspace{-7.5pt}
\caption{Ten measured $I_{leakage-non\_measured\_cells}$ values for each of two CNT-PUF cells corresponding to the logical value of 1 (red and orange) and each of two cells corresponding to the logical value of 0 (blue and green). It is clear that, also in this case, with the drain of the selected cell also being grounded, the nature of the cells cannot be distinguished using these measurements.}
\vspace{-10pt}
\label{fig:idothersIDS}
\end{figure*}

\begin{figure*}[!t]
\centering
\includegraphics[width=0.75\linewidth]{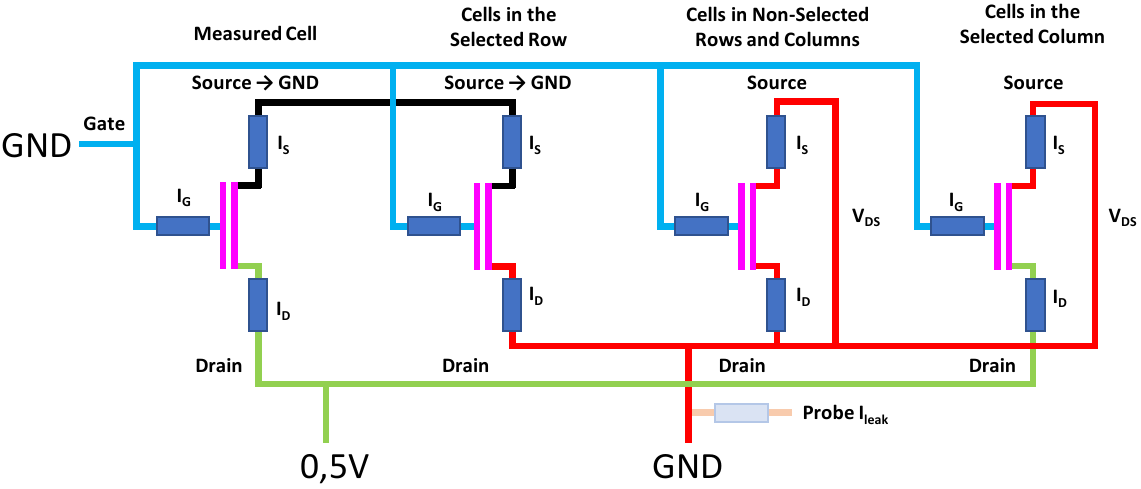}
\vspace{-7.5pt}
\caption{The  equivalent cell layout of the modified second measurement method: Measuring the leakage current of the cells found in rows and columns that have not been selected, with the drains of all cells found in columns other than the selected column and the sources of all cells found in rows other than the selected row being grounded.}
\vspace{-15pt}
\label{fig:twolayoutVDS}
\end{figure*}

\begin{figure*}[!t]
\centering
\includegraphics[width=0.6\linewidth]{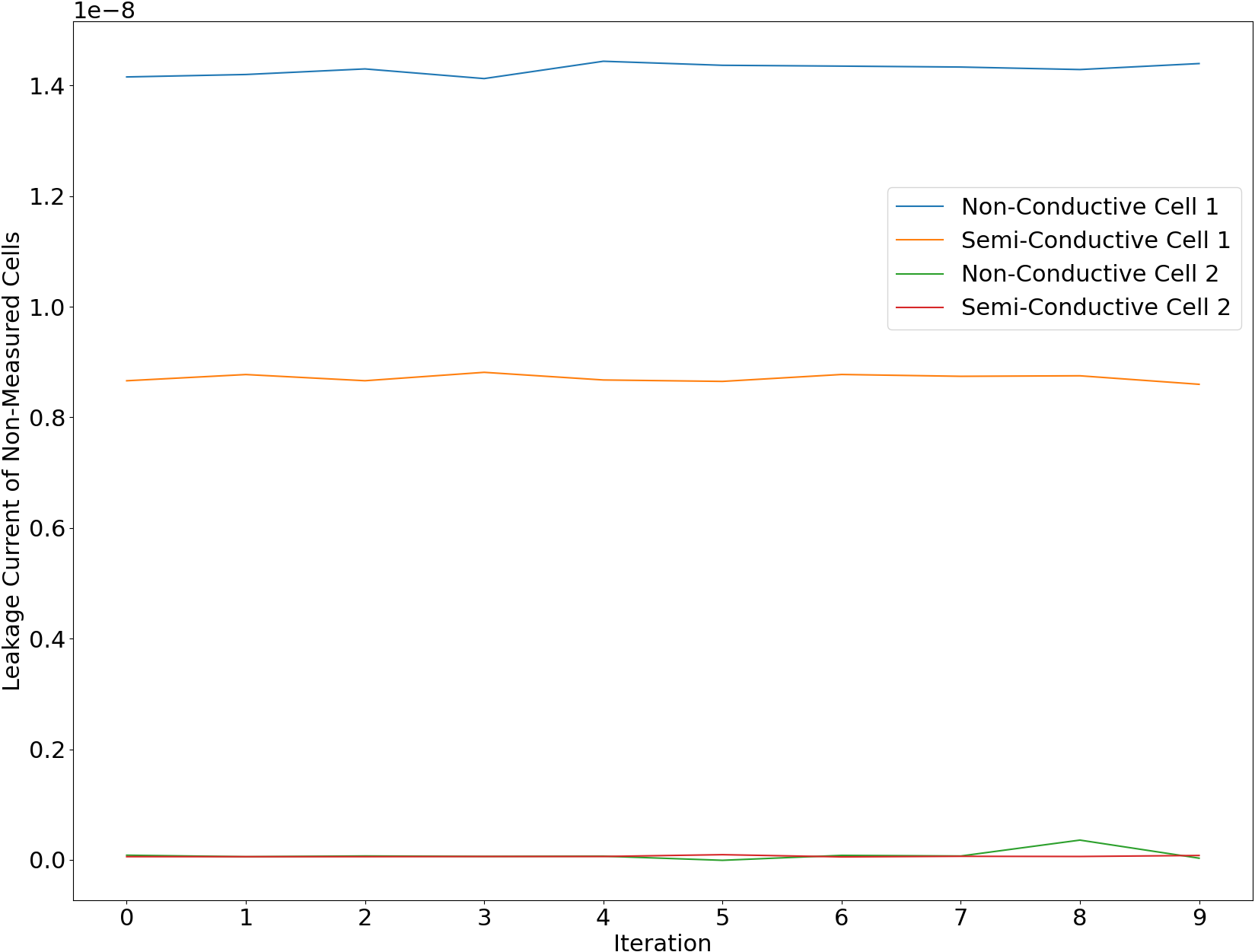}
\vspace{-7.5pt}
\caption{Ten measured $I_{leakage-non\_measured\_cells}$ values for each of two CNT-PUF cells corresponding to the logical value of 1 (red and orange) and each of two cells corresponding to the logical value of 0 (blue and green). It is clear that, also in this case, , with the drains of all cells found in columns other than the selected column and the sources of all cells found in rows other than the selected row being grounded, the nature of the cells cannot be distinguished using these measurements.}
\vspace{-10pt}
\label{fig:idothersVDS}
\end{figure*}

Additionally, similar results occur when either of the following changes is made: the drain of the selected cell is also grounded, or the drains of all cells found in columns other than the selected column and the sources of all cells found in rows other than the selected row are grounded, so that cells found in rows and columns that have both not been selected exhibit $V_{DS}=0$. For the former case, the relevant equivalent cell layout is shown in~\Cref{fig:twolayoutIDS} and the relevant measurement results in~\Cref{fig:idothersIDS}, and for the latter case, the relevant equivalent cell layout is shown in~\Cref{fig:twolayoutVDS} and the relevant measurement results in~\Cref{fig:idothersVDS}.

\subsubsection{Measuring the Drain Leakage current of the Cells Selected}

\begin{figure*}[!t]
\centering
\includegraphics[width=0.75\linewidth]{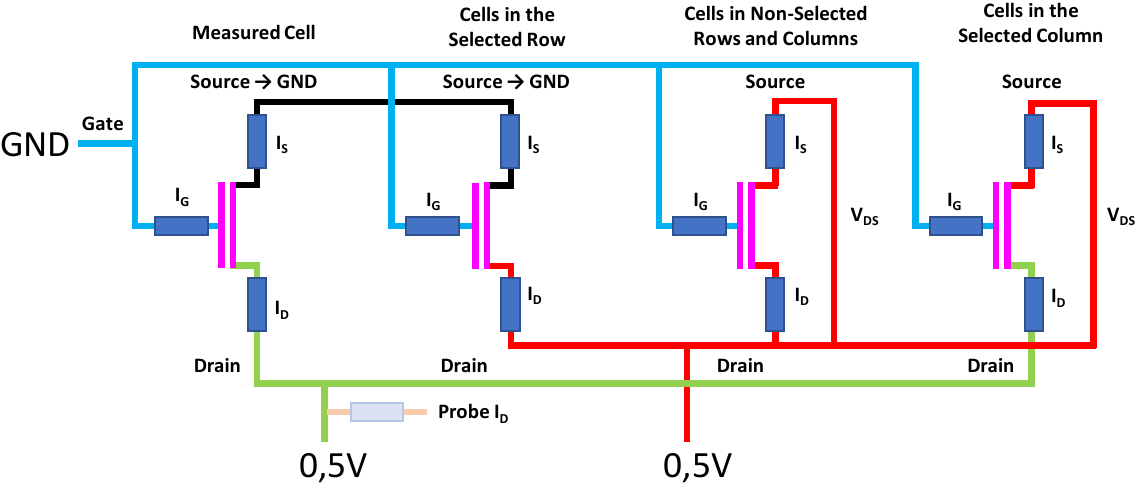}
\vspace{-7.5pt}
\caption{The  equivalent cell layout of the third measurement method: Measuring the drain leakage current of the cells selected.}
\vspace{-15pt}
\label{fig:threelayout}
\end{figure*}

\begin{figure*}[!t]
\centering
\includegraphics[width=0.6\linewidth]{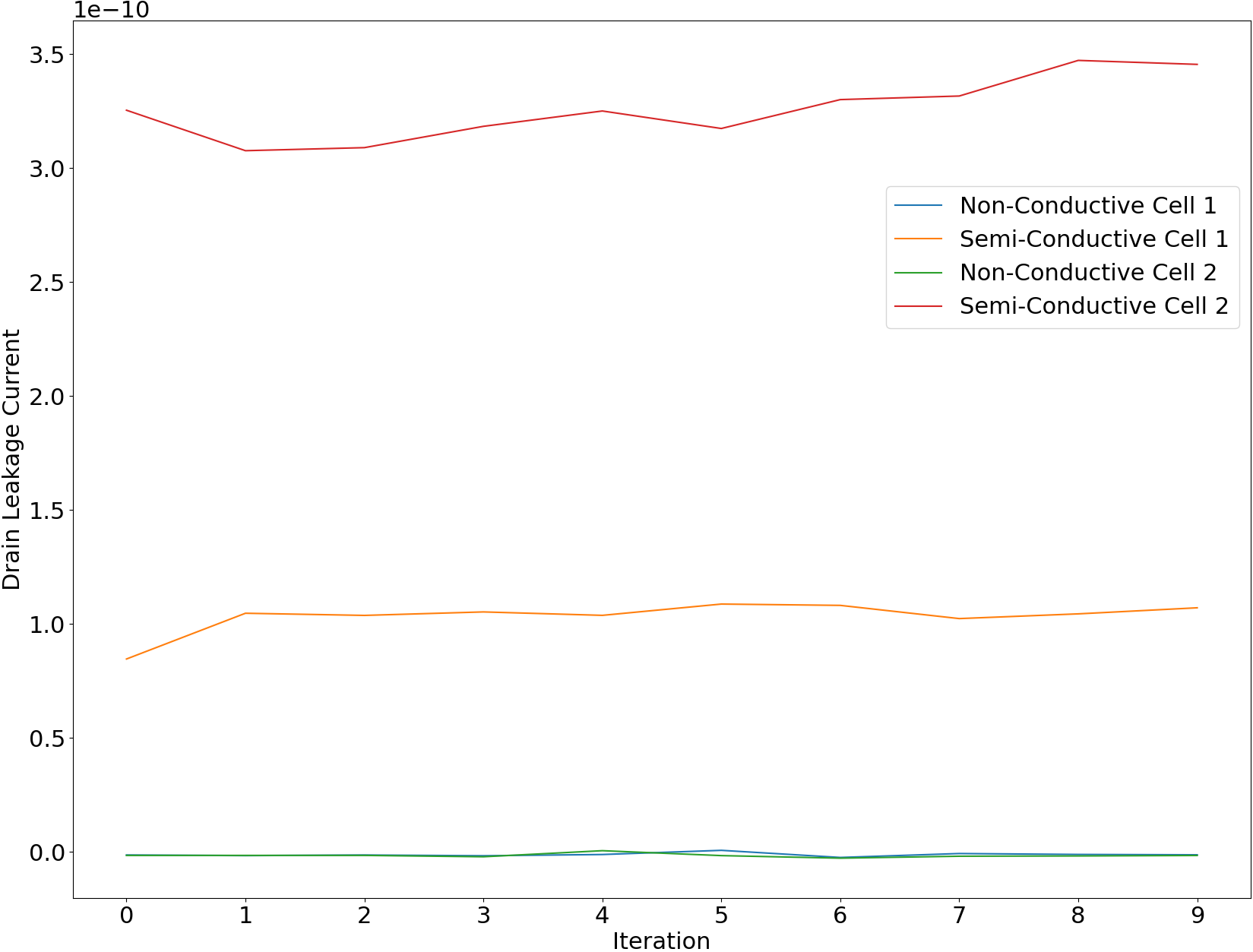}
\vspace{-7.5pt}
\caption{Ten measured $I_{D-leakage}$ values for each of two CNT-PUF cells corresponding to the logical value of 1 (red and orange) and each of two cells corresponding to the logical value of 0 (blue and green). It is clear that, in this case, the nature of the cells can clearly be distinguished using these measurements.}
\label{fig:idleakage}
\vspace{-15pt}
\end{figure*}

We measure the gate leakage current, $I_{D-leakage}$, of cells selected for measurement, by grounding the global gate. Each time, we also ground the selected cell's source and provide $0.5$V to its drain. The sources and the drains of the cells in rows and columns not selected are also provided with $0.5$V, leading these cells to exhibit $V_{DS}=0$. Then, we measure the drain current of the selected cell. As the global gate is grounded, none of the cells is expected to be turned on. The equivalent cell layout for this probing measurement is shown in~\Cref{fig:threelayout}.~\Cref{fig:idleakage} illustrates the results relevant to such probing measurements for two semiconducting and two non-conducting cells~\textsuperscript{\ref{foot3}}. It is evident that, in this case, the nature of the cells can clearly be distinguished using these measurements.

\subsection{Discussion on the Results Presented and Potential Countermeasures}

Our results indicate that non-invasive direct electrical probing can potentially compromise the security of this PUF. At the same time, however, as our results clearly show, the only way for an attacker to gain meaningful information on the nature of a CNT-PUF cell through non-invasive probing is to measure its drain leakage current. Nevertheless, as the drain current of a cell essentially forms its secret, measuring the drain leakage current is rather equivalent to being granted access to the secret itself.

To this end, it is clear that any potential countermeasure should aim to disallow access to the drains of the cells, as these carry the secret. Thus, potential countermeasures may include burying the relevant wire under other wires and metal layers in the Printed Circuit Board (PCB) of a larger system, if the CNT-PUF is directly incorporated into the PCB, or including the relevant characterisation circuits on the CNT-PUF chip, so that there is no need for the drain current to be transmitted over one of the chip's pins, if the CNT-PUF is incorporated into a chip that is soldered to the relevant PCB. 

In any case, we do note that the attacker needs to decide on an adequate threshold in order to distinguish between the drain leakage currents from conductive cells and those from non-conductive cells. This may require some experience, which, we believe, however, that it is easy to gain even during an attack itself. 

Moreover, we believe that the proposed attack technique is rather easy to perform, does not incur extremely high costs, and requires only a relatively small amount of time, in the order of minutes. Thus, the presented attack method can be considered as rather practical. Even if the presented method can be mitigated as an attack, it can, at the same time, serve as an adequate testing method for detecting defective non-conducting CNT-FETs and/or potentially short-circuited cells.

Thus, we can easily conclude that the examined CNT-PUFs are rather resilient to direct probing attacks, and that, in order for the attacker to gain the full-length value of the PUF response, all the relevant drain wires would need to be probed, which could potentially be prevented. At the same time, however, the described non-invasive probing methods appear to be promising for testing such CNT-based structures for the detection of faults.

\section{Conclusion}
\label{sec:conclusion}

In this work, we have presented a number of practical non-invasive attack methods against the CNT-PUF, a novel carbon-nanotube-based physical unclonable function. We have examined these attack techniques and proven that most of them do not provide the attacker with any advantage. At the same time, however, we have shown that by measuring the drain leakage current of each CNT-PUF, the attacker can, with high probability, fully predict the CNT-PUF's response, i.e., its secret.

To this end, and as non-invasive probing attacks are becoming more and more available, we suggest adequately protecting the relevant wires and/or pins that may provide access to a CNT-PUF cell's drain.

In general, we can easily conclude that the examined non-invasive direct probing methods have a higher potential as testing techniques rather than as truly efficient attacks. Nevertheless, even though the examined CNT-PUFs have proven to be rather resilient to direct probing attacks, such attacks are feasible and adequate countermeasures need to be employed in order to address this issue.

As part of future work, it could be examined for the first probing method proposed, if grounding the sources of the cells in non-selected rows, in order to have most of the cells turned on, and only the cells in the selected column, including the cell selected for measurement, turned off, would have any effect on the attacker's ability to distinguish between conducting and non-conducting cells through the gate leakage current.

\section*{Acknowledgment}

This work has been partially funded by the Interreg VI-A Programme Germany/Bavaria--Austria 2021--2027 -- Programm INTERREG VI-A Bayern--Österreich 2021--2027, as part of Project BA0100016: ``CySeReS-KMU: Cyber Security and Resilience in Supply Chains with focus on SMEs'', co-funded by the European Union, and by the German Research Foundation -- Deutsche Forschungsgemeinschaft (DFG), under Projects 440182124: ``PUFMem: Intrinsic Physical Unclonable Functions from Emerging Non-Volatile Memories'', and 439892735: ``NANOSEC: Tamper-Evident PUFs Based on Nanostructures for Secure and Robust Hardware Security Primitives'' of the Priority Program -- SchwerPunktProgramme (SPP) 2253: ``Nano Security: From Nano-Electronics to Secure Systems''.

\bibliographystyle{./IEEEtran}
\bibliography{./IEEEabrv,./IEEEexample}

\end{document}